\documentclass[conference]{IEEEtran}
\IEEEoverridecommandlockouts

\usepackage{cite}
\usepackage{amsmath,amssymb,amsfonts}
\usepackage{algorithmic}
\usepackage{graphicx}
\usepackage{textcomp}
\usepackage{xcolor}

\usepackage{amsfonts}
\usepackage{mathrsfs}
\usepackage{amssymb}   
\usepackage{algorithm}
\usepackage{algorithmic}
\usepackage{cite}      
\usepackage{graphicx}
\usepackage{epstopdf}
\usepackage{psfrag}    
\usepackage{subfigure} 
\usepackage{url}       
\usepackage{stfloats}  
\usepackage{enumerate}
\usepackage{bm}
\usepackage{fancyhdr}
\usepackage{colortbl}
\usepackage{threeparttable}
\usepackage{cases}

\begin{document}

\title{Directional Sparsity Based Statistical Channel Estimation for 6D Movable Antenna Communications}

\author{\IEEEauthorblockN{Xiaodan Shao$\IEEEauthorrefmark{1}$, Rui Zhang$\IEEEauthorrefmark{2}\IEEEauthorrefmark{3}$, Jihong Park$\IEEEauthorrefmark{4}$, Tony Q. S. Quek$\IEEEauthorrefmark{4}$, Robert Schober$\IEEEauthorrefmark{5}$, and Xuemin (Sherman) Shen$\IEEEauthorrefmark{1}$}
	\IEEEauthorblockA{$\IEEEauthorrefmark{1}$Department of Electrical and Computer Engineering, University of Waterloo, Canada\\
		$\IEEEauthorrefmark{2}$School of Science and Engineering, Chinese University of Hong Kong, Shenzhen, China\\
		$\IEEEauthorrefmark{3}$Department of Electrical and Computer Engineering, National University of Singapore, Singapore\\
		$\IEEEauthorrefmark{4}$Information System Technology and Design Pillar, Singapore University of Technology and Design, Singapore\\
		$\IEEEauthorrefmark{5}$Institute for Digital Communications, Friedrich-Alexander University of Erlangen-N\"{u}rnberg, Germany\\
		E-mails: x6shao@uwaterloo.ca, elezhang@nus.edu.sg, jihong\_park@sutd.edu.sg, tonyquek@sutd.edu.sg, \\ robert.schober@fau.de, sshen@uwaterloo.ca}\vspace{-19pt}
}\maketitle

\begin{abstract}
Six-dimensional movable antenna (6DMA) is an innovative and transformative technology to improve wireless network capacity by adjusting the 3D positions and 3D rotations of  antennas/surfaces (sub-arrays) based on the channel spatial distribution. For optimization of the antenna positions and rotations, the acquisition of statistical channel state information (CSI) is essential for 6DMA systems. In this paper, we unveil for the first time a new \textbf{\textit{directional sparsity}}  property of the 6DMA channels between the base station (BS) and the distributed users, where each user has significant channel gains only with a (small) subset of 6DMA position-rotation pairs, which can receive direct/reflected signals from the user. By exploiting this property, a covariance-based algorithm is proposed for estimating the statistical CSI in terms of the average channel power at a small number of 6DMA positions and rotations. Based on such limited channel power estimation, the average channel powers for all possible 6DMA positions and rotations in the BS movement region are reconstructed by further estimating the multi-path average power and direction-of-arrival (DOA) vectors of all users.
Simulation results show that the proposed directional sparsity-based algorithm can achieve higher channel power estimation accuracy than existing benchmark schemes, while requiring a lower pilot overhead. 
\end{abstract}
\section{Introduction}
The forthcoming six-generation (6G) wireless network is expected to bring revolutionary advancements, including improved communication reliability and sensing capabilities \cite{saad, you, shaos}. To achieve these ambitious goals, the trend is to equip the networks with more antennas, evolving from massive multiple-input multiple-output (MIMO) to extremely large-scale MIMO \cite{10045774,lar}. However,
this approach leads to increased hardware cost, energy consumption, and complexity. A key drawback of current MIMO systems is the fixed positioning of antennas at the base station (BS). With fixed-position antennas (FPAs) \cite{lar}, the network cannot adapt its antenna resources to changes in the three-dimensional (3D) spatial distribution of the users.

Recently, to fully exploit the spatial variation of wireless channels
at the transmitter/receiver, six-dimensional movable antenna (6DMA) comprised of multiple rotatable and positionable antennas/sub-arrays  has been proposed as a new technology to improve the performance of wireless networks without increasing the number of antennas \cite{shao20246d, 6dma_dis, shaoaga}. As shown in Fig. \ref{practical_scenario}, equipped with distributed 6DMA surfaces to match the spatial user distribution, a
6DMA-enabled transmitter/receiver can adaptively allocate antenna resources in space to maximize the array gain and spatial multiplexing gain while also suppressing interference. 
In practice, the positions and rotations of 6DMAs can be adjusted in a continuous \cite{shao20246d} or discrete \cite{6dma_dis} manner based on the surface movement mechanism. Channel estimation for 6DMA is investigated in \cite{10883029}. Moreover, 6DMA can also significantly improve sensing accuracy \cite{censing, shao2025tutorial} compared to existing FPAs \cite{lar} and fluid antennas \cite{xiao,10906511}.

Although the benefits of 6DMA in wireless communication and sensing
systems have been demonstrated, the channel state information (CSI) acquisition problem for 6DMA still remains largely unexplored in the literature. To obtain optimal 6DMA positions and rotations, the CSI of the channels between all 6DMA candidate positions/rotations and all users is essential. However, unlike conventional FPAs, which have a finite number of channel coefficients to estimate, 6DMA channels require CSI acquisition in a continuous space with a vast number of candidate antenna positions and/or rotations. Furthermore, existing methods for fluid antennas \cite{xiao} only adjust positions along a finite line or two-dimensional (2D) surface and focus on instantaneous channel estimation, making them inapplicable to CSI estimation for 6DMA. 

To efficiently solve the channel estimation problem for 6DMA
systems, we propose in this paper a new method that exploits 
a unique property of 6DMA channels, which we refer to as \textbf{\textit{directional sparsity}}, i.e., the fact that each user has significant channel gains only for a (small) subset of 6DMA position-rotation pairs that can receive signals of dominant strength given the user's location. By exploiting directional sparsity, the average channel power
is first recovered for a small number of channel measurements taken by the
6DMA at random positions and rotations. Then, the multi-path average power and direction-of-arrival (DOA) vectors of all users are determined based on the recovered average channel power, enabling the reconstruction of the average channel power between the user and all possible 6DMA positions/rotations in the BS movement region. Simulation results show that the
proposed channel reconstruction method is more efficient
than benchmark schemes in terms of pilot overhead and yet
achieves higher channel reconstruction accuracy.

\emph{Notations}: The operator $\mathbb{E}[\cdot]$ denotes the expectation, $\left \|\cdot\right \|_2$, $\left \|\cdot\right \|_F$, and  $\mathbf{I}_N$ denote the Euclidean norm, the Frobenius norm, and the $N \times N$ identity matrix, respectively. $[\mathbf{a}]_j$ and $[\mathbf{A}]_{i,j}$ represent the $j$-th element of a vector $\mathbf{a}$ and the $(i,j)$ element of a matrix $\mathbf{A}$, respectively. For a matrix \(\mathbf{A}\), $\text{tr}(\mathbf{A})$ denotes its trace, $[\mathbf{A}]_{i,:}$ and $[\mathbf{A}]_{:,i}$ denote its $i$-th row and its $i$-th column, respectively, and \(\det(\mathbf{A})\) denotes its determinant.
\section{System Model}
\subsection{6DMA Architecture}
\begin{figure}[t!]
	\centering
	\vspace{-0.69cm}
	\setlength{\abovecaptionskip}{0.cm}
	\includegraphics[width=3.2in]{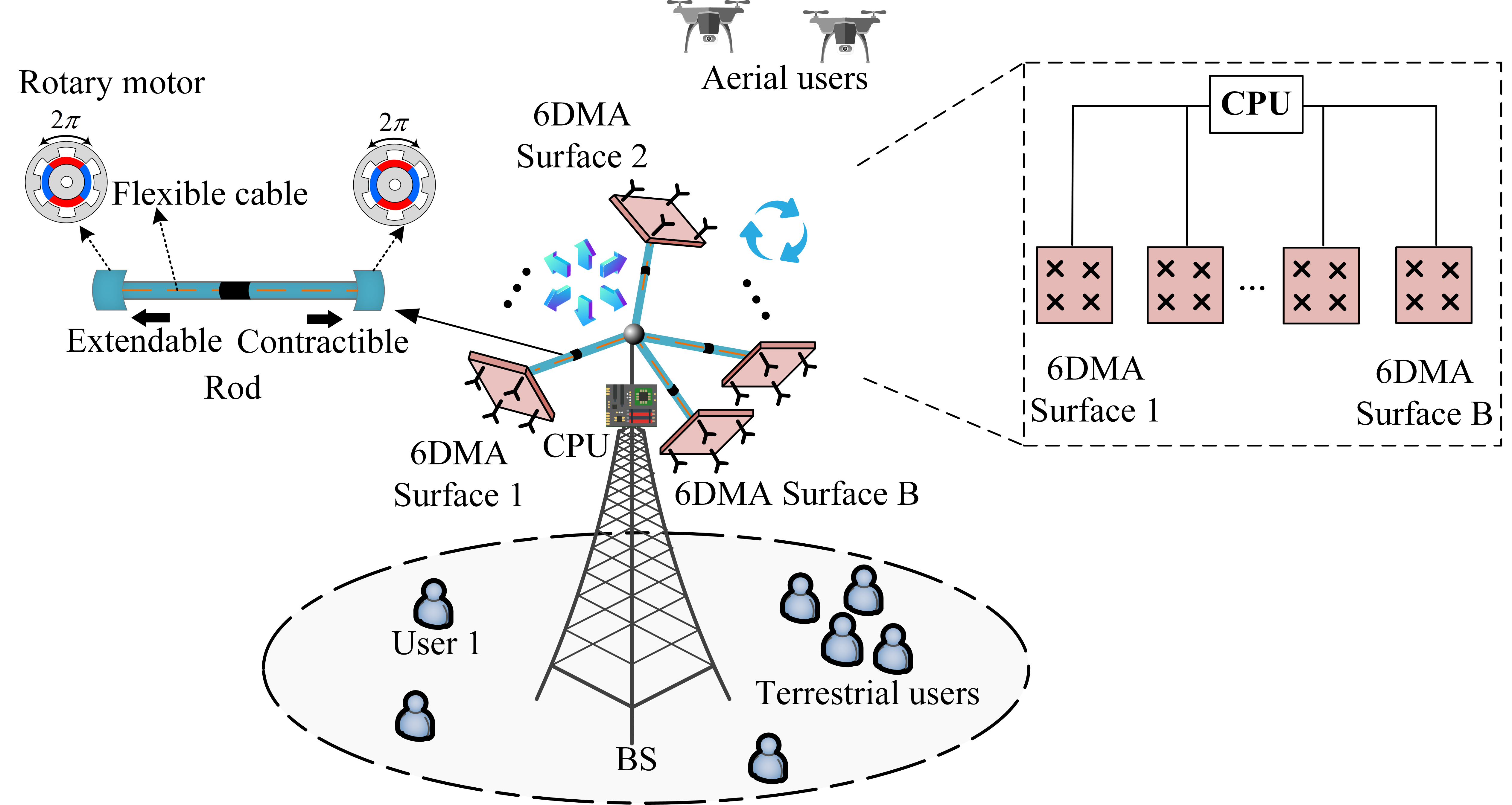}
	\caption{6DMA-equipped BS and signal processing architecture.}
	\label{practical_scenario}
	\vspace{-0.66cm}
\end{figure}
As illustrated in Fig. \ref{practical_scenario}, the proposed 6DMA-BS comprises $B$ distributed 6DMA surfaces, indexed by set $\mathcal{B} = \{1, 2, \ldots, B\}$. Each surface is modeled as a uniform planar array (UPA) with $N \geq 1$ antennas, indexed by set $\mathcal{N} = \{1, 2, \ldots, N\}$. As shown in Fig. \ref{practical_scenario},
each 6DMA surface is connected to a common central processing unit (CPU) at the
BS via a separate rod, which houses flexible cables (e.g., coaxial cables). These cables supply power to the 6DMA surface, and also facilitate control and signal exchange between the 6DMA surface and CPU. Moreover, the CPU carries out the necessary baseband processing tasks and controls two rotary motors that are mounted at both ends of the rod to adjust the position and rotation of each 6DMA surface. In addition, each rod can flexibly contract and
extend to control the distance between each 6DMA surface and
the CPU. For this setup, the BS can function akin to
a ``transformer", possessing the capability to instantly reconfigure its antenna array into virtually any conceivable shape for optimizing wireless network performance. 

Let $\mathbf{q}_b\!=\![x_b,y
_b,z_b]^T\!\in \! \mathbb{R}^3$ and $\mathbf{u}_b\!=\![\alpha_b,\beta_b,\gamma_b]^T\!\in\!\mathbb{R}^3, b\in \mathcal{B}$, denote the position and rotation of the $b$-th 6DMA surface, respectively. 
\(x_b\), \(y_b\), and \(z_b\) are the coordinates of the center of the $b$-th 6DMA surface in a global Cartesian coordinate system (CCS) with origin \(o\), where the CPU is located. All rotation angles \(\alpha_b\), \(\beta_b\), and \(\gamma_b\) lie in \([0, 2\pi)\) and correspond to rotations around the \(x\)-, \(y\)-, and \(z\)-axes, respectively. Let space \(\mathcal{C}\) define the 3D region at the BS site where the 6DMA surfaces can be dynamically positioned and rotated.

The rotation matrix $\mathbf{R}(\mathbf{u}_b)$ for the angles in $\mathbf{u}_b$ is given by
\begin{align}\label{R}
	\scalebox{0.8}{$ %
		\begin{aligned}
			\mathbf{R}(\mathbf{u}_{b}) =
			&\begin{bmatrix}
				c_{\beta_{b}}c_{\gamma_{b}} & c_{\beta_{b}}\omega_{\gamma_{b}} & -\omega_{\beta_{b}} \\
				\omega_{\beta_{b}}\omega_{\alpha_{b}}c_{\gamma_{b}}-c_{\alpha_{b}}
				\omega_{\gamma_{b}} & \omega_{\beta_{b}}\omega_{\alpha_{b}}\omega_{\gamma_{b}}+c_{\alpha_{b}}c_{\gamma_{b}} & c_{\beta_{b}}\omega_{\alpha_{b}} \\
				c_{\alpha_{b}}\omega_{\beta_{b}}c_{\gamma_{b}}+\omega_{\alpha_{b}}\omega_{\gamma_{b}} & c_{\alpha_{b}}\omega_{\beta_{b}}\omega_{\gamma_{b}}-\omega_{\alpha_{b}}c_{\gamma_{b}} &c_{\alpha_{b}}c_{\beta_{b}} \\
			\end{bmatrix},
		\end{aligned}$}
\end{align}
where $c_{x}=\cos(x)$ and $\omega_{x}=\sin(x)$. We set a local CCS $o'\text{-}x'y'z'$ for the 6DMA surface at each location with $o'$ being the center of the surface. The position of the $n$-th antenna on the 6DMA surface in its local CCS is denoted by $\bar{\mathbf{r}}_{n}\in \mathbb{R}^3$. Rotation $\mathbf{u}_b$ and position $\mathbf{q}_b$ are then used to find the position of the $n$-th antenna on the $b$-th 6DMA surface in the global CCS. This is given by $\mathbf{r}_{b,n}\in \mathbb{R}^3$ as follows: 
\begin{align}\label{nwq}
\!\!\!\!\mathbf{r}_{b,n}(\mathbf{q}_b,\mathbf{u}_b)=\mathbf{q}_b+\mathbf{R}
(\mathbf{u}_b)\bar{\mathbf{r}}_{n},~n\in\mathcal{N},~b \in\mathcal{B}.
\end{align}
\subsection{Channel Model}
We focus on uplink multiuser transmission, where \( K \) users, each equipped with a single FPA, are distributed throughout the cell. Assuming a multi-path channel between each user and the BS, the channel from user \( k \) to the \( B \) 6DMA surfaces is defined as $	\mathbf{h}_{k}(\mathbf{q},\mathbf{u})=
[\mathbf{h}_{1,k}^T(\mathbf{q}_{{1}},\mathbf{u}_{{1}}),\cdots, \mathbf{h}_{B,k}^T(\mathbf{q}_{{B}},\mathbf{u}_{{B}})]^T\in \mathbb{C}^{NB}$ with
${\mathbf{q}}=[\mathbf{q}_{1}^T,\cdots,\mathbf{q}_{B}
	^T]^T\in \mathbb{R}^{3B}$ and 
	$\mathbf{u}=[\mathbf{u}_{1}^T,\cdots,\mathbf{u}_{B}^T]^T\in \mathbb{R}^{3B} \label{lk1}$.
Here, $\mathbf{h}_{b,k}(\mathbf{q}_{b},\mathbf{u}_{b})\in \mathbb{C}^{N}$ represents the channel from the $k$-th user to all the antennas of the $b$-th 6DMA surface at the BS, which is given by
\begin{align}\label{uk}
\mathbf{h}_{b,k}(\mathbf{q}_{b},\mathbf{u}_{b})\!=\!
\sum_{\iota=1}^{\Gamma_{k}}\sqrt{\mu_{\iota, k}}e^{-j \varphi_{\iota, k}}
\sqrt{g_{\iota,k}(\mathbf{u}_{b},\mathbf{f}_{\iota,k})}\mathbf{a}_{\iota,k}
(\mathbf{q}_{b},\mathbf{u}_{b}),
\end{align}
where $\Gamma_{k}$ represents the total number of channel paths from user \( k \) to the BS, and ${\mu}_{\iota, k}$ and $\varphi_{\iota, k}$ denote the path gain and phase shift from user \( k \) to the BS along path \( \iota \), respectively.

In \eqref{uk}, the 6D steering vector of the \( b \)-th 6DMA surface for receiving a signal from user \( k \) over path \( \iota \) is expressed as follows: 
\begin{align}\label{gen}
&\mathbf{a}_{\iota,k}(\mathbf{q}_{b},\mathbf{u}_{b})
= \!\left[\!e^{-j\frac{2\pi}{\lambda}
\mathbf{f}_{\iota,k}^T\mathbf{r}_{b,1}(\!\mathbf{q}_{b},
\mathbf{u}_{b}\!)},
\!\cdots,\! e^{-j\frac{2\pi}{\lambda}\mathbf{f}_{\iota,k}^T
\mathbf{r}_{b,N}(\!\mathbf{q}_{b},\mathbf{u}_{b}\!)}\!\right]^T,
\end{align}
for $\iota\in \{1,\ldots,\Gamma_k\}$, where $\lambda$ denotes the carrier wavelength, and $\mathbf{f}_{\iota,k}$ represents the unit-length DOA vector corresponding to direction $(\theta_{\iota,k}, \phi_{\iota,k})$, which is defined as
\begin{align}\label{KM}
\!\!\mathbf{f}_{\iota,k}\!=\![\cos(\theta_{\iota,k})\cos(\phi_{\iota,k}), \cos(\theta_{\iota,k})\sin(\phi_{\iota,k}), \sin(\theta_{\iota,k})]^T.\!
\end{align}

Here, the azimuth and elevation angles for the \(\iota\)-th channel path between user \(k\) and the BS are given by \(\phi_{\iota,k}\in[-\pi,\pi]\) and \(\theta_{\iota,k}\in[-\pi/2,\pi/2]\), respectively. The effective antenna gain of the \(b\)-th 6DMA surface along direction \((\tilde{\theta}_{b,\iota,k}, \tilde{\phi}_{b,\iota,k})\) in the linear scale is defined as follows:
\begin{align}\label{gm}
g_{\iota,k}(\mathbf{u}_{b},\mathbf{f}_{\iota,k})=10^{\frac{A(\tilde{\theta}_{b,\iota,k}, \tilde{\phi}_{b,\iota,k})}{10}},
\end{align}
with
$\tilde{\theta}_{b,\iota,k}=\pi/2-\arccos(\tilde{z}_{b,\iota,k})$, $\tilde{\phi}_{b,\iota,k}=
\arccos\left(\frac{\tilde{x}_{b,\iota,k}}{\sqrt{\tilde{x}_{b,\iota,k}^2+\tilde{y}_{b,\iota,k}^2}}
\right)\times\tau(\tilde{y}_{b,\iota,k})$, 
$[\tilde{x}_{b,\iota,k},\tilde{y}_{b,\iota,k},\tilde{z}_{b,\iota,k}]^T
=-\mathbf{R}(\mathbf{u}_{b})^T\mathbf{f}_{\iota,k}$, $\tau(\tilde{y}_{b,\iota,k})$ returns $1$ if $\tilde{y}_{b,\iota,k} \geq 0$ and $-1$ if $\tilde{y}_{b,\iota,k} < 0$, and \(A(\tilde{\theta}_{b,\iota,k}, \tilde{\phi}_{b,\iota,k})\) denotes the effective antenna gain in dBi determined by the antenna radiation pattern.

\subsection{Achievable Sum Rate Analysis}
Let matrix $
\mathbf{H}(\mathbf{q},\mathbf{u})\!=\![\mathbf{h}_1(\mathbf{q},\mathbf{u}),\cdots,
	\mathbf{h}_{K}(\mathbf{q},\mathbf{u})])\in \mathbb{C}^{BN\times K}$ 
represent the multiple-access channel from all $K$ users to all $B$ 6DMA surfaces at the BS. By  expecting with respect to (w.r.t.) the random channel ${\mathbf{H}}(\mathbf{q},\mathbf{u})$, the achievable ergodic sum rate of the users 
is given by 
\begin{align}\label{qaaa}
\!\!\!\!\!C(\mathbf{q},\mathbf{u})\!=\!\mathbb{E}\left[\log_2 \det \left(\mathbf{I}_{K}+\frac{p}{\sigma^2}\left[{\mathbf{H}}(\mathbf{q},\mathbf{u})^H
{\mathbf{H}}(\mathbf{q},\mathbf{u})\right]\right)\right],
\end{align}
where $\sigma^2$ and $p$ denote average noise power
and the transmit power of each user, respectively.
The exact ergodic sum rate is hard to obtain,
hence we resort to deriving an upper bound for it by exploiting Jensen's inequality. Assuming that the channels of different users are statistically independent, $C(\mathbf{q},\mathbf{u})$ can be upper-bounded by 
\begin{subequations}
\label{10}
\begin{align} 
&\!\!\!\!\!	C(\mathbf{q},\mathbf{u})\leq 
\log_2 \det \left(\mathbf{I}_{K}+\frac{p}{\sigma^2}\mathbb{E}\left[{\mathbf{H}}(\mathbf{q},\mathbf{u})^H
	{\mathbf{H}}(\mathbf{q},\mathbf{u})\right]\right)
\label{o4}\\
	&\!\!\!\!\!=\log_2 \det \!\left(\!\mathbf{I}_{K}+\frac{p}{\sigma^2}\mathbb{E}
	\left[\sum_{j=1}^{NB}[{\mathbf{H}}(\mathbf{q},\mathbf{u})]_{j,:}^H[{\mathbf{H}}(\mathbf{q},\mathbf{u})]_{j,:}\right]\!\right)\!	\label{o3}\\
	&\!\!\!\!\!=\sum_{k=1}^K\log_2 \left( 1+\frac{p}{\sigma^2}\sum_{b=1}^{B}[\mathbf{P}]_{b,k}\right),\label{o1}
\end{align}
\end{subequations}
where \eqref{o1} holds due to the assumed independence of the channels of different users, and
\(\mathbf{P} \in \mathbb{R}^{B \times K}\) represents the average channel power matrix, with its \((b,k)\)-th element, \([\mathbf{P}]_{b,k}\!\!=\!\!\sum_{j=N(b-1)+1}^{Nb} \mathbb{E}\!\left[\!\left|[{\mathbf{H}}(\mathbf{q},\mathbf{u})]_{j,k}\right|^2\right]\), corresponding to the average power of the channels between user \(k\) and all antennas on the $b$-th 6DMA surface.

Note that the ergodic sum rate for 6DMAs in \eqref{qaaa} and its upper bound in \eqref{10} 
depend on the statistical CSI (i.e.,  \(\mathbf{P}\)), and thus, on the antenna position/rotation. Hence, the statistical CSI for all
candidate positions and rotations has to be estimated for optimizing the 6DMA surfaces' position and rotation for maximization of the ergodic sum rate.

\vspace{-3pt}
\section{Proposed Statistical CSI Estimation Scheme for 6DMA}
\vspace{-2pt}
In this section, we present the proposed new statistical channel estimation method to efficiently estimate the average channel power for 6DMAs with affordable complexity. The proposed scheme is based on the directional sparsity property of the 6DMA channels and implemented in two steps. First, all $B$ 6DMA surfaces move over a total of $M\geq B$ different position-rotation pairs to collect channel measurement for estimation of the channel power. Second, based on the estimated channel power, the multi-path average power (defined in Section III-B) and the DOA vector are determined for reconstructing the average channel power for all possible 6DMA positions and rotations. 
\begin{figure}[t!]
	\centering
	\vspace{-0.69cm}
		\setlength{\abovecaptionskip}{-3pt}
\setlength{\belowcaptionskip}{-15pt}
	\includegraphics[width=2.8in]{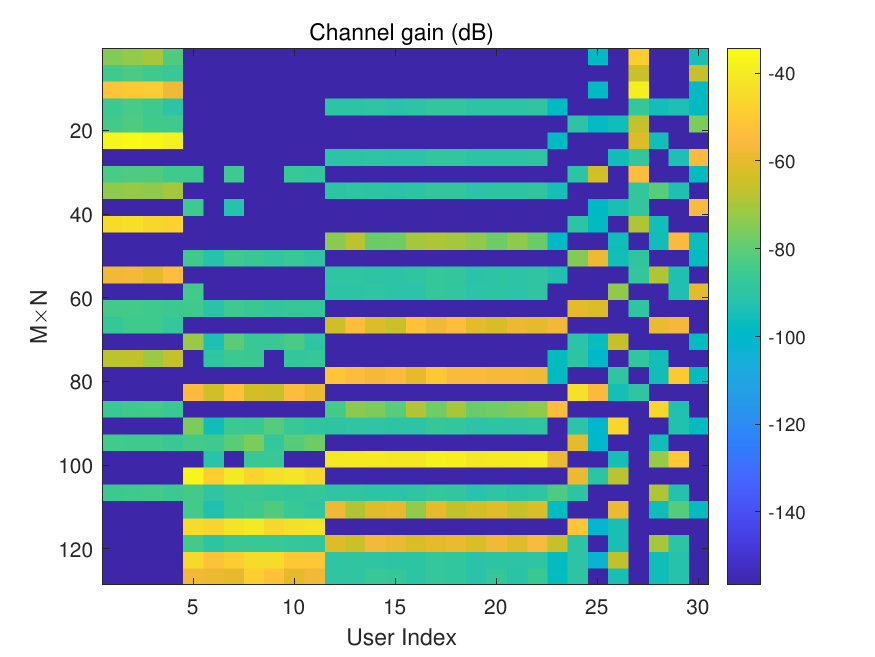}
	\caption{Illustration of the sparsity pattern in $\overline{\mathbf{H}}$ (see Section IV for parameter settings).}
	\label{gain_N}
	\vspace{-0.66cm}
\end{figure}
\vspace{-3pt}
\subsection{Directional Sparsity of 6DMA Channels}
The multiple-access channel from the $K$ users to 6DMA surfaces located at $M$ different position-rotation pairs for channel estimation is characterized by $\overline{\mathbf{H}}=[\overline{\mathbf{H}}_{1}^T,\cdots,\overline{\mathbf{H}}_{M}^T]^T
\in\mathbb{C}^{MN\times K}$ with 
$\overline{\mathbf{H}}_m=
	[\mathbf{h}_{m,1}(\mathbf{q}_{m},\mathbf{u}_{m}),\cdots,\mathbf{h}_{m,K}(\mathbf{q}_{m},\mathbf{u}_{m})]\in \mathbb{C}^{N\times K}$
denoting the channel matrix from all $K$ users to all the antennas of the $m$-th 6DMA candidate position-rotation pair. Here,  $\mathbf{h}_{m,k}(\mathbf{q}_{m},\mathbf{u}_{m})\in \mathbb{C}^{N}$ represents the channel from the $k$-th user to all the antennas of the 6DMA surface at the $m$-th candidate position-rotation pair, and follows a Gaussian distribution with zero mean.

In existing wireless networks, the BS is usually installed at a high altitude. Thus, there are only a limited number of scatterers around the BS, while there may be rich scattering near each of the users. Due to the rotatability, positionability, and antenna directivity of 6DMA, the channels between a given user and different candidate 6DMA positions and rotations in a continuous 3D space generally exhibit drastically different power distributions.
In fact, each user $k$ may have channels with significant gains only for a small subset of all possible 6DMA position-rotation pairs. While for the remaining candidate position-rotation pairs, which may either face in the opposite direction of the user or be blocked by obstacles towards the user, the effective channels of the user are much weaker and thus can be ignored. For
example, in Fig. \ref{practical_scenario}, user 1 can establish a significant channel
with 6DMA surface 1, but its channel with
6DMA surface 2 is much weaker and
thus can be assumed to be approximately zero. We refer to this property as \textbf{\textit{directional sparsity}}.

{\textbf{Definition 1 (Directional Sparsity)}}:
In the considered 6DMA channel model in \eqref{uk}, user $k$ is assumed to have non-zero channel gains to only a subset of 6DMA position-rotation pairs indexed by the set $\mathcal{W}_k\subseteq \mathcal{M}=\{1,2,\cdots,M\}$, 
(i.e., antenna gain $g_{\iota,k}(\mathbf{u}_{m},\mathbf{f}_{\iota,k})\neq 0, \forall m\in \mathcal{W}_k$); while for the remaining 6DMA position-rotation pairs $m\in \mathcal{W}_k^c$, where $
\mathcal{W}_k \cup \mathcal{W}_k^{\mathrm{c}} = \mathcal{M}$ and $ \mathcal{W}_k \cap \mathcal{W}_k^{\mathrm{c}} = \emptyset
$, the channels to user $k$ are assumed to be zero (i.e.,  $g_{\iota,k}(\mathbf{u}_{m},\mathbf{f}_{\iota,k})=0, \forall m\in \mathcal{W}_k^{\mathrm{c}}$). 

The directional sparsity of  \(\overline{\mathbf{H}} \in \mathbb{C}^{MN \times K}\) is illustrated in Fig. \ref{gain_N} \footnote{Note that the directional sparsity applies not only to the channel \(\overline{\mathbf{H}}\) exploited for channel estimation, but also to the 6DMA channel \(\mathbf{H}(\mathbf{q}, \mathbf{u})\) used for data transmission.}. It is interesting to observe that the channel gains of $\overline{\mathbf{H}}$ exhibit a `{\it block sparsity}' pattern because all 
$N$ antennas on a 6DMA surface for a given position-rotation pair share the same channel power distribution. To characterize the directional sparsity of 6DMA channels, we define $\mathbf{Z}\in \mathbb{R}^{M \times K}$ as the directional sparsity indicator matrix as follows: 
 \begin{align}
	[\mathbf{Z}]_{m,k} = \begin{cases} 
		1, & \text{if the channel between the}~ k \text{-th user} \\
		&\text{and the}~ m \text{-th candidate 6DMA}\\ &  \text{position-rotation is non-zero}, \\
		0, & \text{otherwise}.
	\end{cases}
\end{align}

Note that the directional sparsity in the context of 6DMA is different from the traditional concept of angular channel sparsity in MIMO systems with FPAs, which generally refers to the scenario where the number of dominant channel paths of a user is significantly smaller than the number of FPAs at the transmitter/receiver, regardless of their positions/rotations.      
\subsection{Statistical Channel Reconstruction Algorithm}
\subsubsection{Covariance-Based Average Channel Power Estimation}
To enable average channel power estimation in Step I, the 6DMA surfaces at the BS move over $M$ different position-rotation pairs, and the BS collects $L$ pilot symbols from each user. It is 
assumed that the pilots from user $k$, denoted by $\mathbf{x}_k \in \mathbb{C}^{L}$, follow an independent and identically distributed (i.i.d.) complex Gaussian distribution with zero mean and unit variance. Then, the received signal $\mathbf{Y}_m\in \mathbb{C}^{L \times N}$ at the \( m \)-th candidate position-rotation pair can be expressed as
\begin{subequations}
\begin{align}
	\mathbf{Y}_m&=\sum_{k=1}^K[\mathbf{Z}]_{m,k}\mathbf{x}_k \mathbf{h}_{m, k}(\mathbf{q}_{m},\mathbf{u}_{m})^T+\mathbf{W}_m, \label{aps}\\
	&=\mathbf{X}\text{diag}([\mathbf{Z}]_{m,:})\overline{\mathbf{H}}_m^T+\mathbf{W}_m,\label{aps1}
\end{align}
\end{subequations}
where $\mathbf{X}=[\mathbf{x}_1,\cdots,\mathbf{x}_K]\in \mathbb{C}^{L \times K}$ denotes the horizontal stack of all pilots from all $K$ users, and $\mathbf{W}_m \in \mathbb{C}^{L \times N}$ is the noise matrix whose i.i.d. entries follow complex Gaussian distribution $\mathcal{CN}(0,\sigma^2)$. 

Next, we introduce the power state vector \(\boldsymbol{\eta}_m\in\mathbb{R}^{K} \), which is defined as
\begin{align}\label{wq}
\boldsymbol{\eta}_m\!=\!\left[[\overline{\mathbf{P}}]_{m,1}[\mathbf{Z}]_{m,1}, [\overline{\mathbf{P}}]_{m,2}[\mathbf{Z}]_{m,2}, \!\cdots,\! [\overline{\mathbf{P}}]_{m,K}[\mathbf{Z}]_{m,K}\right]^T,\!\!\!
\end{align}
where $\overline{\mathbf{P}}\in \mathbb{R}^{M \times K}$ denotes the statistical CSI (average channel power) for \( M \) position-rotation pairs, with its \((m,k)\)-th element given by \(\sum_{j=N(m-1)+1}^{Nm} \mathbb{E}[|[\overline{\mathbf{H}}]_{j,k}|^2]\).

From \eqref{aps1}, we observe that each column of $\mathbf{Y}_m$, denoted as $[\mathbf{Y}_m]_{:,n}$, $1 \leq n \leq N$, can be modeled as an independent sample drawn from a multivariate complex Gaussian distribution, i.e., 
\begin{equation}\label{ml}
	[\mathbf{Y}_m]_{:,n}\sim \mathcal{CN}(\mathbf{0},\mathbf{X}\text{diag}(\boldsymbol{\eta}_m)\mathbf{X}^H+\sigma^2\mathbf{I}_L).
\end{equation}

Now, we aim to jointly determine 6DMA directional sparsity matrix $\mathbf{Z}$ and estimate average channel power matrix $\overline{\mathbf{P}}$, which then will be used for the statistical CSI reconstruction for all possible positions/rotations.
This involves estimating the power state vector $\boldsymbol{\eta}_m$ from the noisy observations $\mathbf{Y}_m$ by exploiting the known pilot matrix $\mathbf{X}$. In general, this estimation can be formulated as a maximum likelihood estimation (MLE) problem \cite{zhilin}. 
Specifically, we define
$\boldsymbol{\Sigma}_m=\mathbf{X}\text{diag}(\boldsymbol{\eta}_m)\mathbf{X}^H+\sigma^2\mathbf{I}_L$.
Then, the likelihood function of $\mathbf{Y}_m$ given $\boldsymbol{\eta}_m$ can be expressed as 
\begin{align}
	P(\mathbf{Y}_m|\boldsymbol{\eta}_m)
	&=\prod_{n=1}^{N}\frac{1}{\det(\pi\boldsymbol{\Sigma}_m)}
	\exp(-[\mathbf{Y}_{m}]_{:,n}^H\boldsymbol{\Sigma}_m^{-1}[\mathbf{Y}_m]_{:,n})
	\label{ui}\nonumber\\
	&=\frac{1}{\det(\pi\boldsymbol{\Sigma}_m)^N}\exp(-\mathrm{tr}(\boldsymbol{\Sigma}_m^{-1}\mathbf{Y}_m\mathbf{Y}_m^H).
\end{align}

After performing normalization and simplification, we obtain the following equation for the maximum likelihood estimator for $\boldsymbol{\eta}_m$: 
\begin{align}\label{eqr}
\!\!\!\!\!	f(\boldsymbol{\eta}_m)=-\ln P(\mathbf{Y}_m|\boldsymbol{\eta}_m)=\ln\text{det}(\boldsymbol{\Sigma}_m)+\text{tr}(\boldsymbol{\Sigma}_m^{-1}\hat{\boldsymbol{\Sigma}}_{m}),
\end{align}
where $\hat{\boldsymbol{\Sigma}}_{m}=\frac{1}{N}\mathbf{Y}_m\mathbf{Y}_m^H$ denotes the sample covariance matrix of the received signal for the $m$-th candidate position/rotation pair averaged over all $N$ antennas of the 6DMA surface. Based on \eqref{eqr}, the MLE problem can be formulated as 
\begin{align} \label{co}
	\arg \min_{\boldsymbol{\eta}_m\in\mathbb{R}_+}f(\boldsymbol{\eta}_m).
\end{align}
where $\mathbb{R}_+$ denotes the set of all nonnegative real numbers. The solution to \eqref{co} depends on $\mathbf{Y}_m$ through sample covariance matrix $\frac{1}{N}\mathbf{Y}_m\mathbf{Y}_m^H$, whose size scales with $L$
instead of $N$. 
Next, we derive a closed-form expression for the iterative coordinate-wise minimization of \( f(\boldsymbol{\eta}_m) \) in \eqref{eqr}. Let \( k \in \{1,2,\ldots,K\} \) be the index of the  considered coordinate and $\nu$ be the update step. We define \( f_k(\nu) \!=\! f(\boldsymbol{\eta}_m + \nu \mathbf{e}_k) \), where \( \mathbf{e}_k \!\in\! \mathbb{R}^{K} \) is the \( k \)-th canonical basis vector. Then, using the Sherman-Morrison rank-one update identity \cite{sher}, we obtain
\begin{align}\label{sdd}
	\!\!\!\!\!\!&&\!\!\!\!\!\!\left( \boldsymbol{\Sigma}_{m}+\nu\mathbf{x}_k\mathbf{x}_k^H\right )^{-1}=\boldsymbol{\Sigma}_{m}^{-1}-
	\frac{\nu\boldsymbol{\Sigma}_{m}^{-1}\mathbf{x}_k\mathbf{x}_k^H\boldsymbol{\Sigma}_{m}^{-1}}{1+\nu\mathbf{x}_k^H\boldsymbol{\Sigma}_{m}^{-1}\mathbf{x}_k}.
\end{align}
Applying the well-known determinant identity, we have
\begin{align}\label{sdd1}
	\text{det}(\boldsymbol{\Sigma}_{m}+\nu\mathbf{x}_k\mathbf{x}_k^H)=(1+\nu\mathbf{x}_k^H\boldsymbol{\Sigma}_{m}^{-1}\mathbf{x}_k)\text{det}(\boldsymbol{\Sigma}_{m}).
\end{align}
Then, substituting \eqref{sdd} and \eqref{sdd1} into \eqref{eqr} and taking the derivative of $f_k(\nu)$ w.r.t. $\nu$ leads to
\begin{align}\label{gsi}
	\bigtriangledown f_k(\nu)=\frac{\mathbf{x}_k^H\boldsymbol{\Sigma}_{m}^{-1}\mathbf{x}_k}
	{1+\nu\mathbf{x}_k^H\boldsymbol{\Sigma}_{m}^{-1}\mathbf{x}_k}-\frac{\mathbf{x}_k^H\boldsymbol{\Sigma}_{m}^{-1}\hat{\boldsymbol{\Sigma}}_{m}\boldsymbol{\Sigma}_{m}^{-1}\mathbf{x}_k}{(1+\nu\mathbf{x}_k^H\boldsymbol{\Sigma}_{m}^{-1}\mathbf{x}_k)^2}.
\end{align}
The solution of $\bigtriangledown f_k(\nu)=0$ is given by
\begin{align}\label{so}
\bar{\nu}=\frac{\mathbf{x}_k^H\boldsymbol{\Sigma}_{m}^{-1}\hat{\boldsymbol{\Sigma}}_{m}\boldsymbol{\Sigma}_{m}^{-1}\mathbf{x}_k-\mathbf{x}_k^H\boldsymbol{\Sigma}_{m}^{-1}\mathbf{x}_k}{(\mathbf{x}_k^H\boldsymbol{\Sigma}_{m}^{-1}\mathbf{x}_k)^2}.
\end{align}
Then, the elements of $\boldsymbol{\eta}_m$ can be updated as   
\begin{align}
	[\boldsymbol{\eta}_m]_k \to [\boldsymbol{\eta}_m]_k + \nu^*,
\end{align}
where the optimal update step is set as \(\nu^* = \max \left(\bar{\nu}, -[\boldsymbol{\eta}_m]_k \right)\) to preserve the positivity of \([\boldsymbol{\eta}_m]_k\) \cite{zhilin}. 
Once \(\boldsymbol{\eta}_m\) has been obtained, the average channel power matrix can be expressed as \(\overline{\mathbf{P}} = [\boldsymbol{\eta}_1, \ldots, \boldsymbol{\eta}_M]^T\).
Furthermore, we set the directional sparsity matrix \([ \mathbf{Z}]_{m,k} = 1\) if \([\boldsymbol{\eta}_m]_k\) exceeds a given threshold \(\epsilon > 0\); otherwise, \([ \mathbf{Z}]_{m,k} = 0\). The corresponding algorithm is presented in Step I of Algorithm 1.
\subsubsection{Multi-Path Average Power and DOA Vector Estimation}
To express the channel power in a more tractable form, we approximate the element of $\mathbf{h}_{m,k}(\mathbf{q}_{m},
\mathbf{u}_{m})$ in \eqref{uk} as follows:
\begin{align}
	&\!\!\!\![\mathbf{h}_{m,k}(\mathbf{q}_{m},
	\mathbf{u}_{m})]_n\nonumber\\
	&\!\!\!\!\approx \sqrt{g_{k}(\mathbf{u}_{m},\mathbf{f}_{k})}
	e^{-j\frac{2\pi}{\lambda}
		\mathbf{f}_{k}^T\mathbf{r}_{m,n}(\!\mathbf{q}_{m},
		\mathbf{u}_{m}\!)}\sum_{\iota=1}^{\Gamma_{k}}\sqrt{\mu_{\iota, k}}e^{-j \bar{\varphi}_{\iota, k}}
	,\label{ukk1}
\end{align}
where $\bar{\varphi}_{\iota, k}={\varphi}_{\iota, k}+\frac{2\pi}{\lambda}
(\mathbf{f}_{\iota, k}-\mathbf{f}_{k})^T\mathbf{r}_{m,n}(\!\mathbf{q}_{m},
\mathbf{u}_{m})$, which is modeled as an independent and uniformly distributed random variable in \([0, 2\pi)\).
Here, \eqref{ukk1} is obtained by assuming that $g_{\iota,k}(\mathbf{u}_{m},\mathbf{f}_{\iota,k})$ remains constant for all $\iota$, i.e., $g_{\iota,k}(\mathbf{u}_{m},\mathbf{f}_{\iota,k}) = g_{k}(\mathbf{u}_{m},\mathbf{f}_{k}), \forall \iota \in \{1, 2, \cdots, \Gamma_k\}$, with $\mathbf{f}_{k}$ denoting the unit DOA vector corresponding to the signal arriving at the BS from the center of the scattering cluster of user $k$.	
Then, the elements of \(\overline{\mathbf{P}}\) can be rewritten as follows:
\begin{subequations}
\begin{align}
	[\overline{\mathbf{P}}]_{m,k}&=
	\sum_{j=N(m-1)+1}^{Nm} \mathbb{E}\left[\left|[\overline{\mathbf{H}}_m]_{j,k}\right|^2\right],\\
	&=N \mathbb{E}\left[\left|[\mathbf{h}_{m,k}(\mathbf{q}_{m},
	\mathbf{u}_{m})]_n\right|^2\right], \forall n\in \mathcal{N}\label{po1}\\
	&=N g_{k}(\mathbf{u}_{m},\mathbf{f}_{k}) s_k, \label{mm}
\end{align}
\end{subequations}
where $s_k=\mathbb{E}[| \sum_{\iota=1}^{\Gamma_{k}}\sqrt{\mu_{\iota, k}}e^{-j \bar{\varphi}_{\iota, k}} |^2]$ represents the multi-path average power from the $k$-th user to the 6DMA-BS, and \eqref{mm} holds because 
$\mathbb{E}[ (\sum_{\iota=1}^{\Gamma_{k}}\!\sqrt{\mu_{i, k}}e^{-j \bar{\varphi}_{i, k}})(\sum_{\iota=1}^{\Gamma_{k}}\!\sqrt{\mu_{j, k}}e^{-j \bar{\varphi}_{j, k}})]\!=\!0$ for $i\!\neq \!j$.

From \eqref{mm}, we know that the 6DMA channel power is a function of the multi-path average power \(s_k\) and the unit-length DOA vector \(\mathbf{f}_{k}\). Therefore, by determining \(\mathbf{f}_{k}\) and \(s_k\) based on the estimated $\overline{\mathbf{P}}$ and $\mathbf{Z}$ in Step I, the average channel power between users and all possible 6DMA positions and rotations in the BS movement region can be reconstructed.
In particular, we leverage the knowledge of channel directional sparsity matrix $\mathbf{Z}$ estimated in Step I to reduce the parameter estimation complexity in Step II.
Specifically, we construct the support set \(\hat{\mathcal{I}}_k \in \mathbb{R}^{M_k}\) for \([\mathbf{Z}]_{:,k}\), which includes the indices of non-zero elements of \([\mathbf{Z}]_{:,k}\), with \(M_k < M\) denoting the number of the non-zero elements.
We define $\bar{\mathbf{p}}_k=[\overline{\mathbf{P}}]_{:,k}\in \mathbb{R}^{M}$ and $\mathbf{v}_k =[ g_{k}(\mathbf{u}_{1},\mathbf{f}_{k}),\cdots, g_{k}(\mathbf{u}_{M},\mathbf{f}_{k})]^T\in \mathbb{R}^{M}$. 
Then, let \(\bar{\mathbf{p}}_{k,\hat{\mathcal{I}}_k} \in \mathbb{R}^{M_k}\) and \(\mathbf{v}_{k,\hat{\mathcal{I}}_k} \in \mathbb{R}^{M_k}\) denote the new vectors formed by the non-zero elements of \(\bar{\mathbf{p}}_k\) and \(\mathbf{v}_k\) that correspond to the indices stored in \(\hat{\mathcal{I}}_k\). 
Consequently, we optimize 
$\mathbf{f}_{k}$ and $s_k$ by minimizing the reconstruction error:
\begin{subequations} \label{ac}
	\begin{align}
		~&~ \min_{ s_k, \mathbf{f}_{k} } \left\|\bar{\mathbf{p}}_{k,\hat{\mathcal{I}}_k} - N\mathbf{v}_{k,\hat{\mathcal{I}}_k} s_k  \right\|^2_2,\\
		~&~ \text{s.t.}~~~s_{k}\ge 0.
	\end{align}
\end{subequations}

To efficiently estimate $\mathbf{f}_{k}$ and $s_k$ by compressed sensing, we
approximate $\mathbf{f}_{k}$ by uniformly discretizing it into $G$ grid points with $G\geqslant 1$. Thus, we have $\mathbf{v}_{k,\hat{\mathcal{I}}_k} s_k  \approx \tilde{\mathbf{V}}_k \tilde{\mathbf{s}}_k$, where $\tilde{\mathbf{V}}_k\in\mathbb{R}^{M_k\times G}$
is an over-complete matrix and $\tilde{\mathbf{s}}_k\in\mathbb{R}^{G}$
is a sparse vector with one non-zero element corresponding to ${s}_k$.
Then, problem \eqref{ac} reduces to
\begin{subequations}
	\label{eg3}
\begin{align}
~&~ \arg \min_{\tilde{\mathbf{s}}_k} \| \bar{\mathbf{p}}_{k,\hat{\mathcal{I}}_k} - N\mathbf{\tilde{V}}_k \tilde{\mathbf{s}}_k   \|_2 \\
	~&~ \mathrm{s.t.} ~ \|\tilde{\mathbf{s}}_k\|_0 = 1, \\
	~&~~~~~~ \tilde{\mathbf{s}}_k \succeq \bf{0}.
\end{align}
\end{subequations}
Problem \eqref{eg3} is a compressed sensing problem that can be solved using classical compressed sensing algorithms, such as non-negative orthogonal matching pursuit (OMP) \cite{let}, to estimate $\tilde{\mathbf{s}}_k$ and obtain the estimated $\hat{s}_k$. Subsequently, the estimated $\hat{\mathbf{f}}_k$, corresponding to the columns of $\tilde{\mathbf{V}}_k$ with non-zero coefficients in $\tilde{\mathbf{s}}_k$, can be obtained accordingly.

Finally, according to \eqref{mm}, the estimate of \({\mathbf{P}} \in \mathbb{R}^{B \times K}\) in \eqref{o1} can be expressed as
$
[\hat{\mathbf{P}}]_{b,k}
=N g_{k}(\mathbf{u}_{b},\hat{\mathbf{f}}_{k}) \hat{s}_k$.

The details of Step II of the proposed algorithm are summarized in Algorithm 1. The complexity of the covariance-based channel power estimation in Step I is $\mathcal{O}(L^2KM)$. The complexity of solving problem \eqref{eg3} is \(\mathcal{O}(M_kG)\), which is lower than the \(\mathcal{O}(MG)\) complexity when directional sparsity is not exploited. Thus, the overall complexity of Algorithm 1 is $\mathcal{O}(L^2KM+M_kG)$.
\begin{algorithm}[t!]
		\caption{Directional Sparsity-Based Statistical Channel Reconstruction Algorithm}
	\label{alg1}
		\begin{algorithmic}[1]
		\STATE \textbf{Input}: $\{\mathbf{Y}_m\}_{m=1}^{M}$, $\mathbf{X}$, $\{\hat{\boldsymbol{\Sigma}}_{m}=\frac{1}{N}\mathbf{Y}_m\mathbf{Y}_m^H\}_{m=1}^{M}$, and number of iterations $T$.
		\STATE {\textit{Step I (Covariance-Based Average Channel Power Estimation)}}
		\STATE \textbf{Initialization}: $\{\boldsymbol{\eta}_m=\mathbf{0}\}_{m=1}^{M}$, $\mathbf{Z}=\mathbf{0}$, $\{\boldsymbol{\Sigma}_m=\sigma^2\mathbf{I}_L\}_{m=1}^{M}$. \\
		\FOR{$t=1 : T$} 
		\FOR{$m=1 : M$} 
		\STATE {Select an index \( k \in \{1,2, \cdots, K\} \) corresponding to the \( k \)-th element of $\boldsymbol{\eta}_m$ randomly;}
		\STATE  Set $\nu^*=\max\left\{\frac{\mathbf{x}_k^H\boldsymbol{\Sigma}_{m}^{-1}\hat{\boldsymbol{\Sigma}}_{m}\boldsymbol{\Sigma}_{m}^{-1}\mathbf{x}_k-\mathbf{x}_k^H\boldsymbol{\Sigma}_{m}^{-1}\mathbf{x}_k}{(\mathbf{x}_k^H\boldsymbol{\Sigma}_{m}^{-1}\mathbf{x}_k)^2},-[\boldsymbol{\eta}_m]_k\right\}$;
		\STATE  Update $[\boldsymbol{\eta}_m]_k=[\boldsymbol{\eta}_m]_k+\nu^*$;
		\STATE Update $\boldsymbol{\Sigma}_m=\boldsymbol{\Sigma}_m+\nu^*\mathbf{x}_k\mathbf{x}_k^H$;
		\ENDFOR
		\ENDFOR
			\STATE Set directional sparsity matrix \( [\mathbf{Z}]_{m,k} = 1 \) if \( [\boldsymbol{\eta}_m]_k \) exceeds a given threshold $ \epsilon >0$.
		\STATE Obtain the average channel power matrix $\overline{\mathbf{P}}=[\boldsymbol{\eta}_1, \cdots, \boldsymbol{\eta}_M]^T\in \mathbb{R}^{M \times K}$.
		\STATE {\textit{Step II (Multi-Path Average Power and DOA Vector Estimation)}}
		\STATE Determine $\hat{s}_k$ and $\hat{\mathbf{f}}_{k}$ for $k=1,2,\cdots,K$ by solving \eqref{eg3}.
		\STATE Obtain $\hat{\mathbf{P}}$ at the BS via $
		[\hat{\mathbf{P}}]_{b,k}
		=N g_{k}(\mathbf{u}_{b},\hat{\mathbf{f}}_{k}) \hat{s}_k$.
		\STATE \textbf{Output}: $\hat{\mathbf{P}}$.
	\end{algorithmic}
\end{algorithm}

\section{Simulation Results}
In the simulation, we set \( N = 4 \), meaning each 6DMA surface is equipped with a \( 2 \times 2 \) UPA with antenna elements spaced by \( \lambda/2 \). We set \( B = 16 \), $M=32$, $\Gamma_{k}=20, \forall k$, $G=500$, $K=50$, and \( \lambda = 0.125 \) m. The multi-path channels per user are generated by firstly randomly generating the user's location in the BS's coverage area (assumed to be a spherical annulus region $\mathcal{L}$ with radial distances from 30 m to 200 m from the BS center) and then generating random scatterers uniformly within a sphere centered at the user's location and with a radius equal to 3 m. $\mathcal{L}$ consists of three distinct hotspot sub-regions, $\mathcal{L}_v$, $v=1,2,3$, and a regular user sub-region $\mathcal{L}_0$, such that $\mathcal{L}=\mathcal{L}_0\cup(\cup_{v=1}^3\mathcal{L}_v)$. The hotspot sub-regions $\mathcal{L}_1$, $\mathcal{L}_2$, and $\mathcal{L}_3$ are 3D spheres centered at 100 m, 60 m, and 40 m from the CPU, with radii of 15 m, 10 m, and 5 m, respectively. Users are distributed in these areas according to a homogeneous Poisson point process, with the proportion of regular users (i.e., background users with low density) to the total number of users given by 0.3. $A(\tilde{\theta}_{b,\iota,k}, \tilde{\phi}_{b,\iota,k})$ in \eqref{gm} is set as the half-space directive antenna pattern \cite{shao20246d}. 

To evaluate the channel reconstruction performance, we assume that \(\overline{M} = 350 \gg M\) candidate position-rotation pairs are evenly distributed across the largest spherical surface with a radius of 1 meter that fits within the 6DMA-BS site space \(\mathcal{C}\). Let $\mathbf{P}_{\mathrm{g}}\in\mathbb{R}^{\overline{M}\times K}$
and $\hat{\mathbf{P}}_{\mathrm{g}}\in\mathbb{R}^{\overline{M}\times K}$ denote the channel power matrix from all users to the $\overline{M}$ receive position-rotation pairs and the estimate of $\mathbf{P}_{\mathrm{g}}$, respectively.
The performance of the proposed average channel power estimation scheme
is evaluated based on the normalized mean square error (NMSE), 
which is defined as 
$\text{NMSE}=\mathbb{E}\left(\frac{\|
		\mathbf{P}_{\mathrm{g}}-
		\hat{\mathbf{P}}_{\mathrm{g}}\|_F^2}{\|\mathbf{P}_{\mathrm{g}}\|_F^2}\right)$. 
		
We compare the proposed algorithm with three baseline schemes. 1) Covariance-based exhaustive measurement: all $\overline{M} \times K$ channel powers of $\mathbf{P}_{\mathrm{g}}$ are directly measured using the covariance-based algorithm in Step I, without applying the parameter estimation in Step II. 2) 
Approximate message passing (AMP) algorithm \cite{vamp}. 3) Block orthogonal matching pursuit (BOMP) algorithm \cite{bomp}. For AMP and BOMP, Step I estimates the row-sparse channel \(\widetilde{{\mathbf{H}}}_m=\text{diag}([\mathbf{Z}]_{m,:})\overline{\mathbf{H}}_m^T \) from \( \mathbf{Y}_m \), and the channel power can then be determined based on this estimate. Step II remains the same as in the proposed algorithm. 
\begin{figure}[t!]
		\vspace{-0.79cm}
	\centering
	\setlength{\abovecaptionskip}{0.cm}
	\includegraphics[width=2.9in]{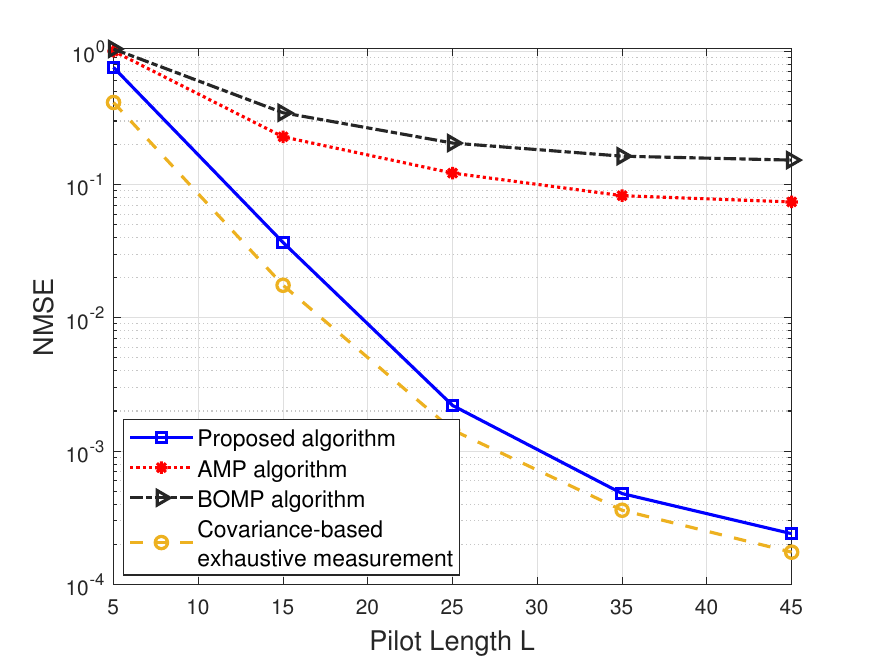}
	\caption{The NMSE of channel power estimation versus pilot length.}
	\label{NMSE_power}
	\vspace{-0.65cm}
\end{figure}

In Fig. \ref{NMSE_power}, we compare the proposed algorithm with the baseline schemes as the length of the pilot sequence increases, with the signal-to-noise ratio (SNR) set to 30 dB.
From this figure, we 
observe that the NMSEs of all considered
algorithms decrease as the pilot length increases, and the proposed algorithm reduces the 
length of the pilot sequence required for achieving a desired accuracy
for the channel reconstruction compared with the AMP and BOMP
benchmark schemes. This improvement is due to the fact that the benchmark schemes have to estimate a much larger number of unknowns to recover the exact entries of \(\widetilde{{\mathbf{H}}}_m\). In addition, the proposed statistical channel reconstruction algorithm has a slightly higher NMSE compared to the covariance-based exhaustive measurement method due to the reconstruction error in Step II. 
However, it requires only $M=32$ sampling 
position-rotation pairs, which is much less than the covariance-based exhaustive measurement method that uses $\overline{M}=350$.  

\begin{figure}[t!]
	\centering
	\vspace{-0.79cm}
	\setlength{\abovecaptionskip}{0.cm}
	\includegraphics[width=2.9in]{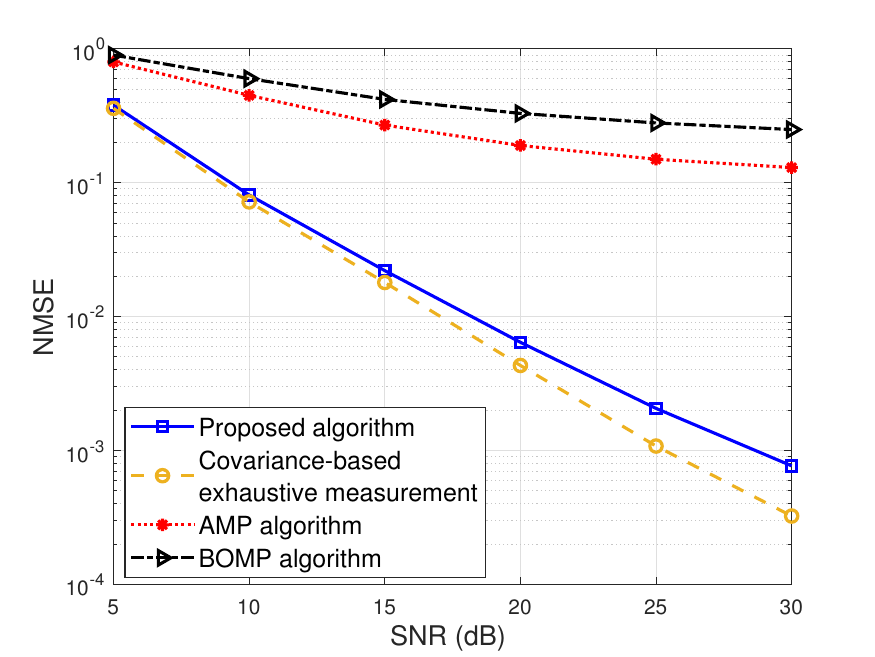}
	\caption{The NMSE of channel power estimation versus SNR.}
	\label{snr}
	\vspace{-0.65cm}
\end{figure}

Fig. \ref{snr} shows the NMSE versus the SNR for
all considered algorithms. We observe that, within the considered SNR range, the proposed algorithm performs better than the AMP and BOMP algorithms but worse than the covariance-based exhaustive measurement method, with the performance gap among them enlarging as the SNR increases. However, to achieve improved performance in the exhaustive measurement method, more position-rotation pairs must be sampled.

\vspace{-3pt}
\section{Conclusions}
\vspace{-3pt}
In this paper, we proposed a new statistical channel reconstruction
method for 6DMA communication systems, which exploits their unique channel  directional sparsity. To reduce the pilot overhead
and computational complexity for channel estimation, the average channel power
was first recovered based on a small number of channel measurements taken at random 6DMA positions and rotations. Then, the multi-path average power and DOA vectors of all users 
were determined based on the channel power estimates to reconstruct the average channel power for all possible 6DMA positions and rotations.
Our simulation results confirm that the proposed channel estimation algorithm achieves higher estimation accuracy than existing schemes while reducing the pilot overhead required.

\vspace{-5pt}
\bibliographystyle{IEEEtran}
\bibliography{fabs}
\end{document}